\documentclass[sigconf,table,bookmarks=false]{acmart}

\usepackage{color}
\usepackage{marvosym}
\usepackage{listings}
\usepackage{amsmath}
\usepackage{graphicx}
\usepackage{cellspace}
\usepackage{xspace}
\usepackage{fancybox}

\newcommand{\tabincell}[2]{\begin{tabular}{@{}#1@{}}#2\end{tabular}}

\newcommand\cincludegraphics[2][]{\raisebox{-0.3\height}{\includegraphics[#1]{#2}}}
\newcommand{\website}{\textcolor{blue}{\url{https://github.com/chenjshnn/LabelDroid}}}
\newcommand{\oursite}{\href{https://github.com/chenjshnn/LabelDroid}{\textcolor{blue}{our website}}}

\newcommand{\jieshan}[1]{\textcolor{blue}{#1}}
\newcommand{\tool}{\textsc{LabelDroid}\xspace}

\AtBeginDocument{%
  \providecommand\BibTeX{{%
    \normalfont B\kern-0.5em{\scshape i\kern-0.25em b}\kern-0.8em\TeX}}}

\copyrightyear{2020}
\acmYear{2020}
\setcopyright{acmcopyright}
\acmConference[ICSE '20]{42nd International Conference on Software Engineering}{May 23--29, 2020}{Seoul, Republic of Korea}
\acmBooktitle{42nd International Conference on Software Engineering (ICSE '20), May 23--29, 2020, Seoul, Republic of Korea}
\acmPrice{15.00}
\acmDOI{10.1145/3377811.3380327}
\acmISBN{978-1-4503-7121-6/20/05}

\acmSubmissionID{icse20-main-1044}

\begin{document}

\title{Unblind Your Apps: Predicting Natural-Language Labels for Mobile GUI Components by Deep Learning}

\author{Jieshan Chen}
\email{Jieshan.Chen@anu.edu.au}
\affiliation{%
  \institution{Australian National University}
  \country{Australia}
  }

\author{Chunyang Chen}
\authornote{Corresponding author.}
\email{Chunyang.Chen@monash.edu}
\affiliation{%
  \institution{Monash University}
  \country{Australia}
}

\author{Zhenchang Xing}
\email{Zhenchang.Xing@anu.edu.au}
\affiliation{%
  \institution{Australian National University}
  \country{Australia}
  }
\additionalaffiliation{%
    \institution{Data61, CSIRO}
    \country{Australia}
}

\author{Xiwei Xu}
\email{Xiwei.Xu@data61.csiro.au}
\affiliation{%
  \institution{Data61, CSIRO}
  \country{Australia}
}

\author{Liming Zhu}
\email{Liming.Zhu@data61.csiro.au}
\affiliation{%
  \institution{Data61, CSIRO}
  \country{Australia}
}
\additionalaffiliation{%
    \institution{University of New South Wales}
    \country{Australia}
}

\author{Guoqiang Li}
\authornotemark[1]
\email{Li.G@sjtu.edu.cn}
\affiliation{%
  \institution{Shanghai Jiao Tong University}
  \country{China}
}
\author{Jinshui Wang}
\authornotemark[1]
\email{ymkscom@gmail.com}
\affiliation{%
  \institution{Fujian University of Technology}
  \country{China}
}

\renewcommand{\shortauthors}{Jieshan Chen, Chunyang Chen, Zhenchang Xing, et al.}

\begin{abstract}
  According to the World Health Organization(WHO), it is estimated that approximately 1.3 billion people live with some forms of vision impairment globally, of whom 36 million are blind.
  Due to their disability, engaging these minority into the society is a challenging problem.
  The recent rise of smart mobile phones provides a new solution by enabling blind users' convenient access to the information and service for understanding the world.
  Users with vision impairment can adopt the screen reader embedded in the mobile operating systems to read the content of each screen within the app, and use gestures to interact with the phone.
  However, the prerequisite of using screen readers is that developers have to add natural-language labels to the image-based components when they are developing the app.
  Unfortunately, more than 77\% apps have issues of missing labels, according to our analysis of 10,408 Android apps.
  Most of these issues are caused by developers' lack of awareness and knowledge in considering the minority.
  And even if developers want to add the labels to UI components, they may not come up with concise and clear description as most of them are of no visual issues.
  To overcome these challenges, we develop a deep-learning based model, called \tool, to automatically predict the labels of image-based buttons by learning from large-scale commercial apps in Google Play.
  The experimental results show that our model can make accurate predictions and the generated labels are of higher quality than that from real Android developers.
\end{abstract}




\maketitle

\section{Introduction}

\begin{figure}
	\centering
	\includegraphics[width=0.48\textwidth]{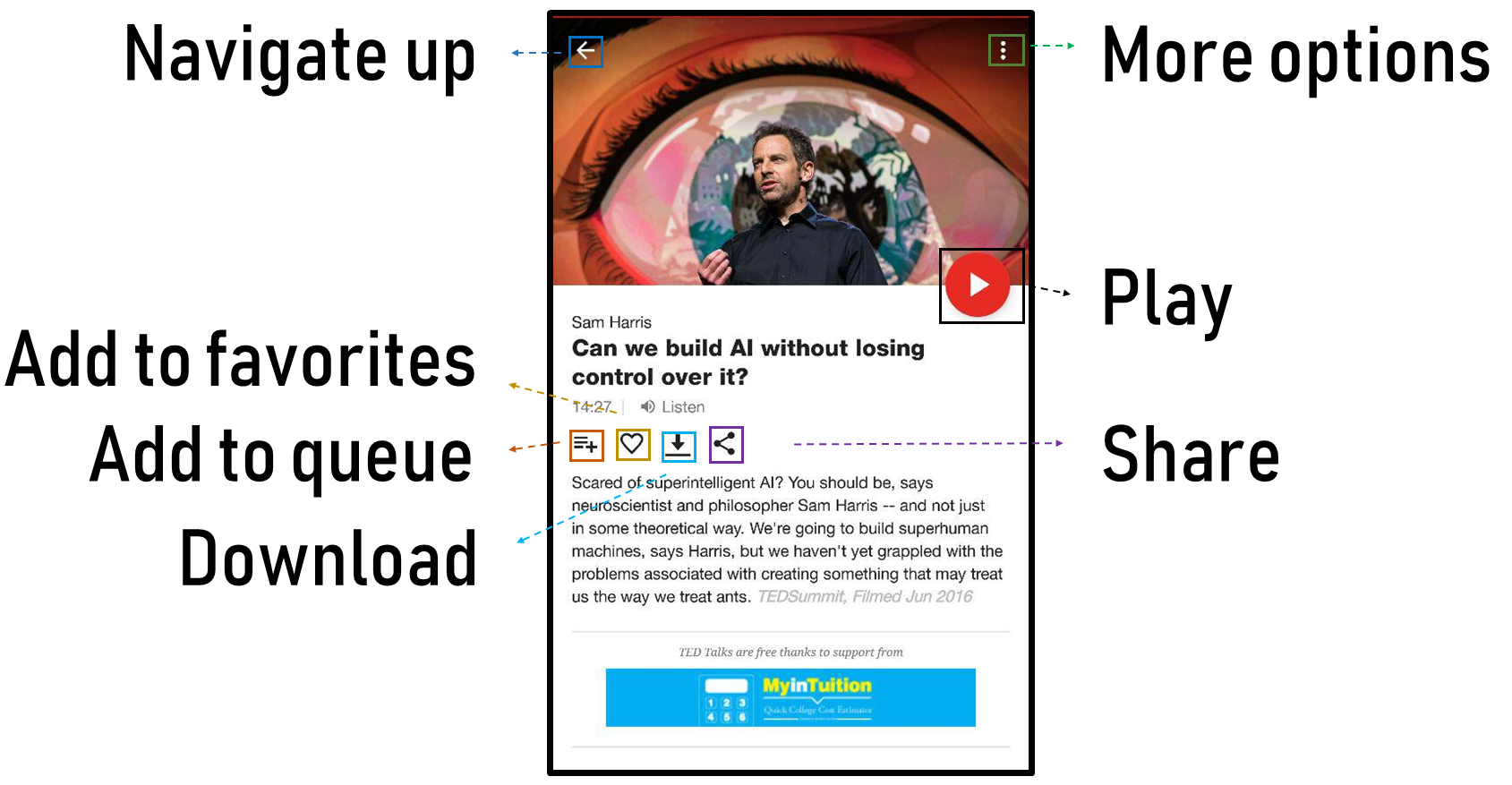}
	\caption{Example of UI components and labels.}
	\label{fig:illustration}
\end{figure}

Given millions of mobile apps in Google Play~\cite{web:googleplay} and App store~\cite{web:appstore}, the smart phones are playing increasingly important roles in daily life.
They are conveniently used to access a wide variety of services such as reading, shopping, chatting, etc.
Unfortunately, many apps remain difficult or impossible to access for people with disabilities.
For example, a well-designed user interface (UI) in Figure~\ref{fig:illustration} often has elements that don't require an explicit label to indicate their purpose to the user.
A checkbox next to an item in a task list application has a fairly obvious purpose for normal users, as does a trash can in a file manager application.
However, to users with vision impairment, especially for the blind, other UI cues are needed.
According to the World Health Organization(WHO)~\cite{web:WHO}, it is estimated that approximately 1.3 billion people live with some form of vision impairment globally, of whom 36 million are blind.
Compared with the normal users, they may be more eager to use the mobile apps to enrich their lives, as they need those apps to represent their eyes.
Ensuring full access to the wealth of information and services provided by mobile apps is a matter of social justice~\cite{ladner2015design}.

Fortunately, the mobile platforms have begun to support app accessibility by screen readers (e.g., TalkBack in Android~\cite{web:talkback} and VoiceOver in IOS~\cite{web:voiceover}) for users with vision impairment to interact with apps.
Once developers add labels to UI elements in their apps, the UI can be read out loud to the user by those screen readers.
For example, when browsing the screen in Figure~\ref{fig:illustration} by screen reader, users will hear the clickable options such as ``navigate up'', ``play'', ``add to queue'', etc. for interaction.
The screen readers also allow users to explore the view using gestures, while also audibly describing what's on the screen.
This is useful for people with vision impairments who cannot see the screen well enough to understand what is there, or select what they need to.

Despite the usefulness of screen readers for accessibility, there is a prerequisite for them functioning well, i.e., the existence of labels for the UI components within the apps.
In detail, the Android Accessibility Developer Checklist guides developers to ``\textit{provide content descriptions for UI components that do not have visible text}"~\cite{web:androidAccessibityGuideline}.
Without such content descriptions\footnote{``labels'' and ``content description'' refer to the same meaning and we use them interchangeably in this paper}, the Android TalkBack screen reader cannot provide meaningful feedback to a user to interact with the app.

\begin{figure}
	\centering
	\includegraphics[width=0.48\textwidth]{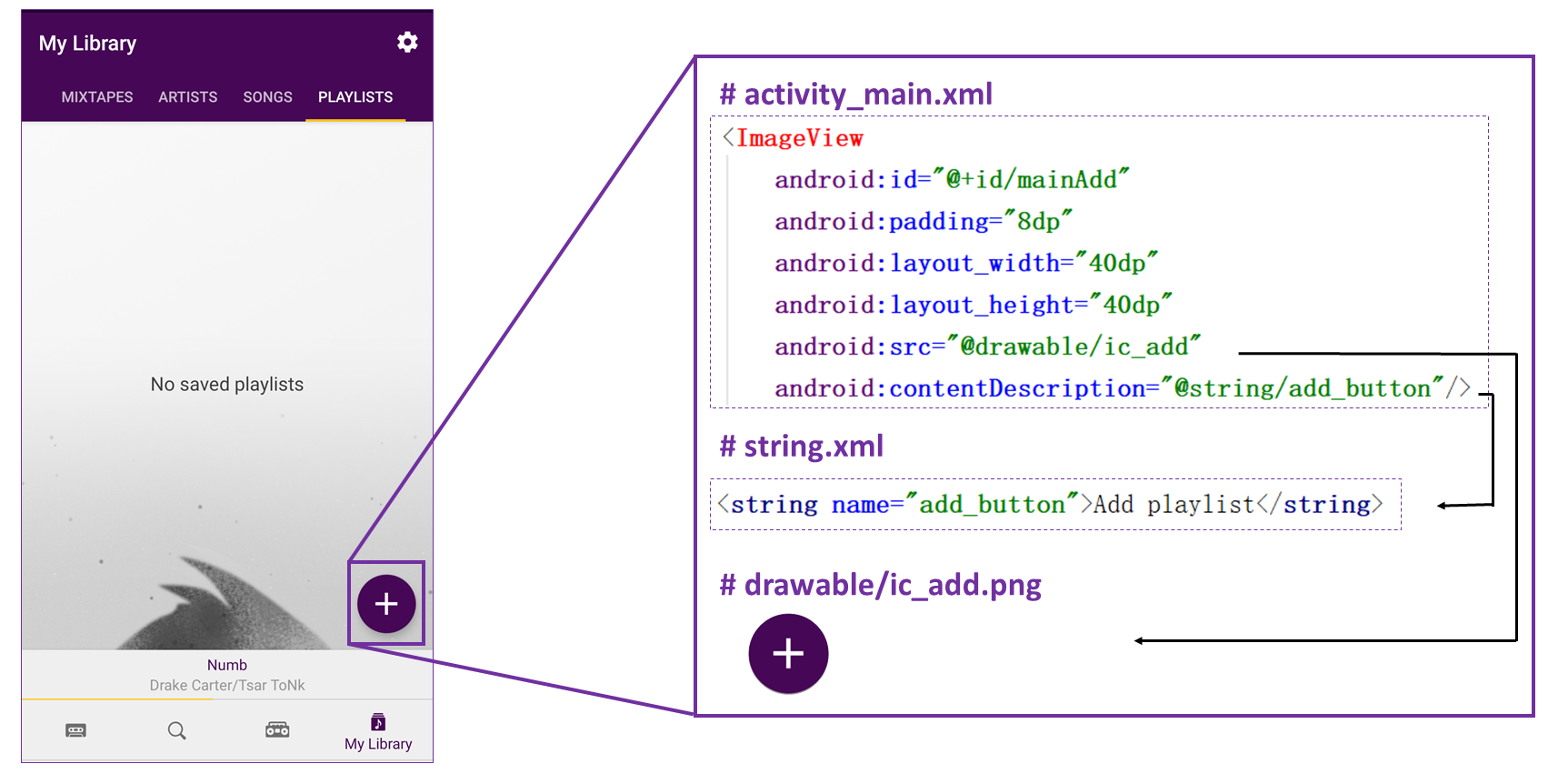}
	\caption{ Source code for setting up labels for ``add playlist'' button (which is indeed a \textit{clickable ImageView}).
	}
	\label{fig:source_code}
\end{figure}

Although individual apps can improve their accessibility in many ways, the most fundamental principle is adhering to platform accessibility guidelines~\cite{web:androidAccessibityGuideline}.
However, according to our empirical study in Section~\ref{sec:motivation}, more than 77\% apps out of 10,408 apps miss such labels for the image-based buttons, resulting in the blind's inaccessibility to the apps.
Considering that most app designers and developers are of no vision issues, they may lack awareness or knowledge of those guidelines targeting for blind users.
To assist developers with spotting those accessibility issues, many practical tools are developed such as Android Lint~\cite{web:androidLint}, Accessibility Scanner~\cite{web:accessibilityScanner}, and other accessibility testing frameworks~\cite{web:androidEspresso, web:androidRobolectric}.
However, none of these tools can help fix the label-missing issues.
Even if developers or designers can locate these issues, they may still not be aware how to add concise, easy-to-understand descriptions to the GUI components for users with vision impairment.
For example, many developers may add label ``add'' to the button in Figure~\ref{fig:source_code} rather than ``add playlist'' which is more informative about the action.

To overcome those challenges, we develop a deep learning based model to automatically predict the content description.
Note that we only target at the image-based buttons in this work as these buttons are important proxies for users to interact with apps, and cannot be read directly by the screen reader without labels.
Given the UI image-based components, our model can understand its semantics based on the collected big data, and return the possible label to components missing content descriptions.
We believe that it can not only assist developers in efficiently filling in the content description of UI components when developing the app, but also enable users with vision impairment to access to mobile apps.

Inspired by image captioning, we adopt the CNN and transformer encoder decoder for predicting the labels based on the large-scale dataset.
The experiments show that our \tool can achieve 60.7\% exact match and 0.654 ROUGE-L score which outperforms both state-of-the-art baselines.
We also demonstrate that the predictions from our model is of higher quality than that from junior Android developers.
The experimental results and feedbacks from these developers confirm the effectiveness of our \tool.

Our contributions can be summarized as follow:
\begin{itemize}
	\item To our best knowledge, this is the first work to automatically predict the label of UI components for supporting app accessibility. We hope this work can invoke the community attention in maintaining the accessibility of mobile apps.
	\item We carry out a motivational empirical study for investigating how well the current apps support the accessibility for users with vision impairment.
	\item We construct a large-scale dataset of high-quality content descriptions for existing UI components. We release the dataset\footnote{\website} for enabling other researchers' future research.
\end{itemize}

\section{Android Accessibility Background}

\subsection{Content Description of UI component}
Android UI is composed of many different components to achieve the interaction between the app and the users.
For each component of Android UI, there is an attribute called \texttt{\small android:contentDescription} in which the natural-language string can be added by the developers for illustrating the functionality of this component.
This need is parallel to the need for alt-text for images on the web.
To add the label ``add playlist'' to the button in Figure~\ref{fig:source_code}, developers usually add a reference to the \textit{android:contentDescription} in the layout xml file, and that reference is referred to the resource file \textit{string.xml} which saves all text used in the application.
Note that the content description will not be displayed in the screen, but can only be read by the screen reader.

\subsection{Screen Reader}
According to the screen reader user survey by WebAIM in 2017~\cite{web:screenreadersurvey7}, 90.9\% of respondents who were blind or visually impaired used screen readers on a smart phone.
As the two largest organisations that facilitate mobile technology and the app market, Google and Apple provide the screen reader (TalkBack in Android~\cite{web:talkback} and VoiceOver in IOS~\cite{web:voiceover}) for users with vision impairment to access to the mobile apps.
As VoiceOver is similar to TalkBack and this work studies Android apps, we only introduce the TalkBack.
TalkBack is an accessibility service for Android phones and tablets which helps blind and vision-impaired users interact with their devices more easily.
It is a system application and comes pre-installed on most Android devices.
By using TalkBack, a user can use gestures, such as swiping left to right, on the screen to traverse the components shown on the screen.
When one component is in focus, there is an audible description given by reading text or the content description.
When you move your finger around the screen, TalkBack reacts, reading out blocks of text or telling you when you've found a button.
Apart from reading the screen, TalkBack also allows you to interact with the mobile apps with different gestures.
For example, users can quickly return to the home page by drawing lines from bottom to top and then from right to left using fingers without the need to locate the home button.
TalkBack also provides local and global context menus, which respectively enable users to define the type of next focused items (e.g., characters, headings or links) and to change global setting.

\begin{figure}
	\centering
    \includegraphics[width=0.4\textwidth]{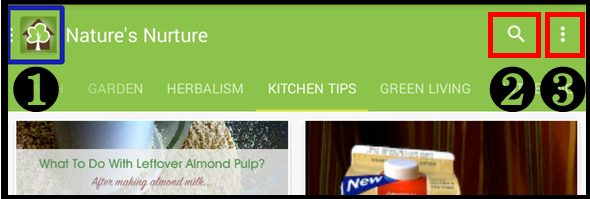}
	\caption{Examples of image-based buttons. \textcircled{1} \texttt{\small clickable ImageView}; \textcircled{2}\textcircled{3} \texttt{\small ImageButton}.
    }
	\label{fig:example}
\end{figure}

\subsection{Android Classes of Image-Based Buttons}
There are many different types of UI components~\cite{web:androidWidget} when developers are developing the UI such as \texttt{\small TextView, ImageView, EditText, Button}, etc.
According to our observation, the image-based nature of the classes of buttons make them necessary to be added with natural-language annotations for two reasons.
First, these components can be clicked i.e., as important proxies for interaction than static components like \texttt{\small TextView} with users.
Second, the screen readers cannot directly read the image to natural language.
In order to properly label an image-based button such that it interacts properly with screen readers, alternative text descriptions must be added in the button's content description field.
Figure~\ref{fig:example} shows two kinds of image-based buttons.

\subsubsection{Clickable Images}
Images can be rendered in an app using the Android API class \texttt{\small android.widget.ImageView}\cite{web:imageview}.
If the clickable property is set to true, the image functions as a button (\textcircled{1} in Figure~\ref{fig:example}).
We call such components Clickable Images.
Different from normal images of which the clickable property is set to false, all Clickable Images are non-decorative, and a null string label can be regarded as a missing label accessibility barrier.

\subsubsection{Image Button}
Image Buttons are implemented by the Android API class \texttt{\small android.widget.ImageButton}~\cite{web:imagebutton}.
This is a sub-class of the Clickable Image's \texttt{\small ImageView} class.
As its name suggests, Image Buttons are buttons that visually present an image rather than text (\textcircled{2}\textcircled{3} in Figure~\ref{fig:example}).

\section{Motivational Mining Study}
\label{sec:motivation}

While the main focus and contribution of this work is developing a model for predicting content description of image-based buttons, we still carry out an empirical study to understand whether the development team adds labels to the image-based buttons during their app development.
The current status of image-based button labeling is also the motivation of this study.
But note that this motivational study just aims to provide an initial analysis towards developers supporting users with vision impairment, and a more comprehensive empirical study would be needed to deeply understand it.

\subsection{Data Collection}
To investigate how well the apps support the users with vision impairment, we randomly crawl 19,127 apps from Google Play~\cite{web:googleplay}, belonging to 25 categories with the installation number ranging from 1K to 100M.

We adopt the app explorer~\cite{chen2018ui} to automatically explore different screens in the app by various actions (e.g., click, edit, scroll).
During the exploration, we take the screenshot of the app GUI, and also dump the run-time front-end code which identifies each element's type (e.g., TextView, Button), coordinates in the screenshots, content description, and other metadata.
Note that our explorer can only successfully collect GUIs in 15,087 apps.
After removing all duplicates by checking the screenshots, we finally collect 394,489 GUI screenshots from 15,087 apps.
Within the collected data, 278,234(70.53\%) screenshots from 10,408 apps contain image-based buttons including clickable images and image buttons, which forms the dataset we analyse in this study.

\begin{table}
	\centering
	\caption{Statistics of label missing situation}
	\resizebox{0.48\textwidth}{!}{
		\begin{tabular}{llll}
			\hline
			\textbf{Element}         &\textbf{\#Miss/\#Apps} & \textbf{\#Miss/\#Screens} & \textbf{\#Miss/\#Elements}   \\
			\hline
			\textbf{ImageButton}     & 4,843/7,814 (61.98\%)  & 98,427/219,302(44.88\%)  & 241,236/423,172(57.01\%)  \\
			\textbf{Clickable Image} & 5,497/7,421 (74.07\%)  & 92,491/139,831(66.14\%)  & 305,012/397,790(76.68\%)  \\
			\hline
			\textbf{Total}           & 8,054/10,408 (77.38\%) & 169,149/278,234(60.79\%) & 546,248/820,962(66.54\%)  \\
			\hline
	\end{tabular}}
	\label{tab:missing_label}
\end{table}

\subsection{Current Status of Image-Based Button Labeling in Android Apps}
Table~\ref{tab:missing_label} shows that 4,843 out of 7,814 apps (61.98\%) have image buttons without labels, and 5,497 out of 7,421 apps (74.07\%) have clickable images without labels.
Among all 278,234 screens, 169,149(60.79\%) of them including 57.01\% image buttons and 76.68\% clickable images within these apps have at least one element without explicit labels.
It means that more than half of image-based buttons have no labels.
These statistics confirm the severity of the button labeling issues which may significantly hinder the app accessibility to users with vision impairment.

\begin{figure}
	\centering
	\includegraphics[width=0.48\textwidth]{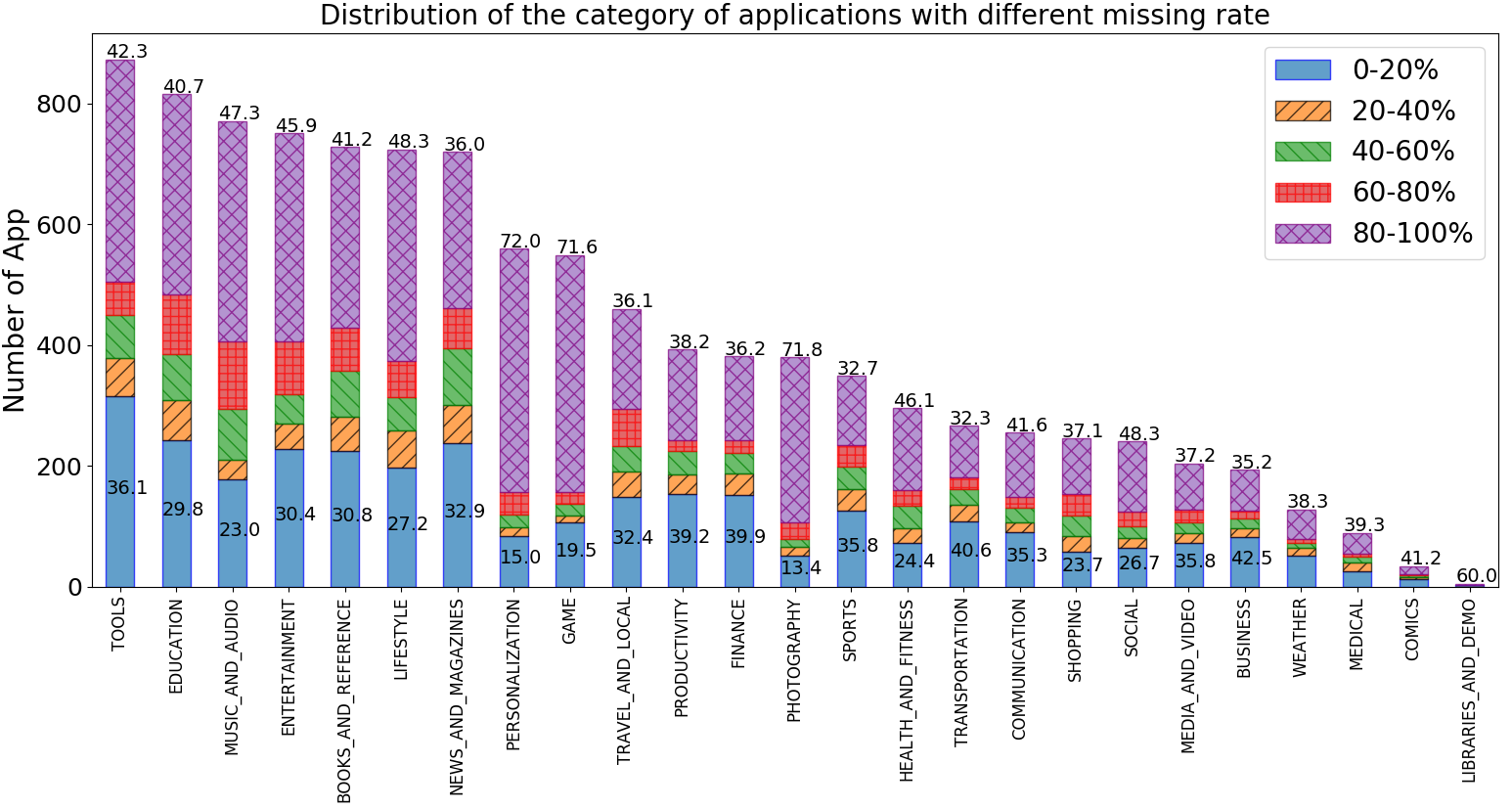}
	\caption{The distribution of the category of applications with different rate of image-based buttons missing content description}
	\label{fig:empirical_study}
\end{figure}

We then further analyze the button labeling issues for different categories of mobile apps.
As seen in Figure~\ref{fig:empirical_study}, the button labeling issues exist widely across different app categories, but some categories have higher percentage of label missing buttons.
For example, 72\% apps in \textit{Personalization}, 71.6\% apps in \textit{Game}, and 71.8\% apps in \textit{Photography} have more than 80\% image-based button without labels.
\textit{Personalization} and \textit{Photography} apps are mostly about updating the screen background, the alignment of app icons, viewing the pictures which are mainly for visual comfort.
The \textit{Game} apps always need both the screen viewing and the instantaneous interactions which are rather challenging for visual-impaired users.
That is why developers rarely consider to add labels to image-based buttons within these apps, although these minority users deserve the right for the entertainment.
In contrast, about 40\% apps in \textit{Finance}, \textit{Business}, \textit{Transportation}, \textit{Productivity} category have relatively more complete labels for image-based buttons with only less than 20\% label missing.
The reason accounting for that phenomenon may be that the extensive usage of these apps within blind users invokes developers' attention, though further improvement is needed.

\begin{figure}
	\centering
	\includegraphics[width=0.48\textwidth]{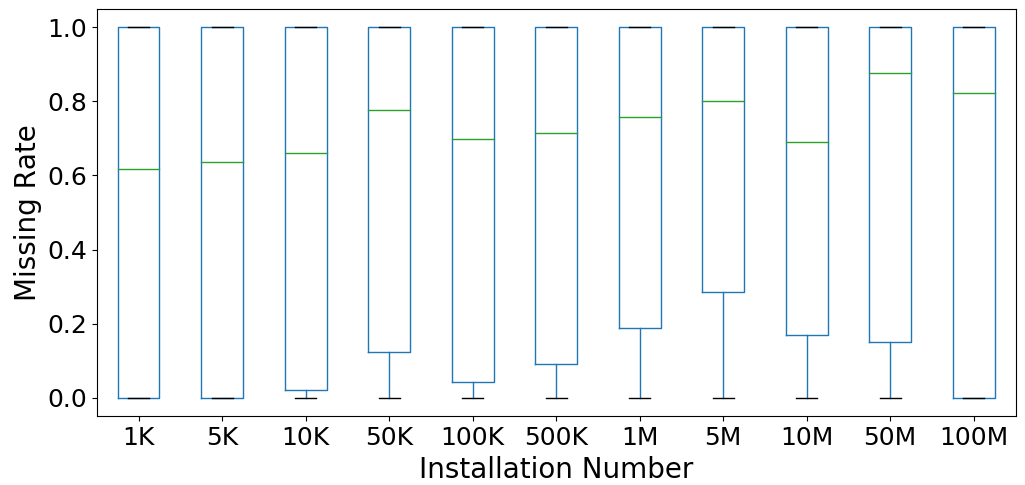}
	\caption{Box-plot for missing rate distribution of all apps with different installation numbers. }
	\label{fig:overall_miss}
\end{figure}

To explore if the popular apps have better performance in adding labels to image-based buttons, we draw a box plot of label missing rate for apps with different installation numbers in Figure~\ref{fig:overall_miss}.
There are 11 different ranges of installation number according to the statistics in Google Play.
However, out of our expectation, many buttons in popular apps still lack the labels.
Such issues for popular apps may post more negatively influence as those apps have a larger group of audience.
We also conduct a Spearman rank-order correlation test~\cite{spearman2010proof} between the app installation number and the label-missing rate.
The correlation coefficient is 0.046 showing a very weak relationship between these two factors.
This result further proves that the accessibility issue is a common problem regardless of the popularity of applications.
Therefore, it is worth developing a new method to solve this problem.


\noindent\fbox{
	\parbox{0.95\linewidth}{
		\textbf{Summary}: By analyzing 10,408 existing apps crawled from Google Play, we find that more than 77\% of them have at least one image-based button missing labels.
		Such phenomenon is further exacerbated for apps categories highly related to pictures such as personalization, game, photography .
		However, out of our expectation, the popular apps do not behave better in accessibility than that of unpopular ones.
		These findings confirm the severity of label missing issues, and motivate our model development for automatic predicting the labels for image-based buttons.
	}
}

\begin{figure*}
	\centering
    \includegraphics[width=0.95\textwidth]{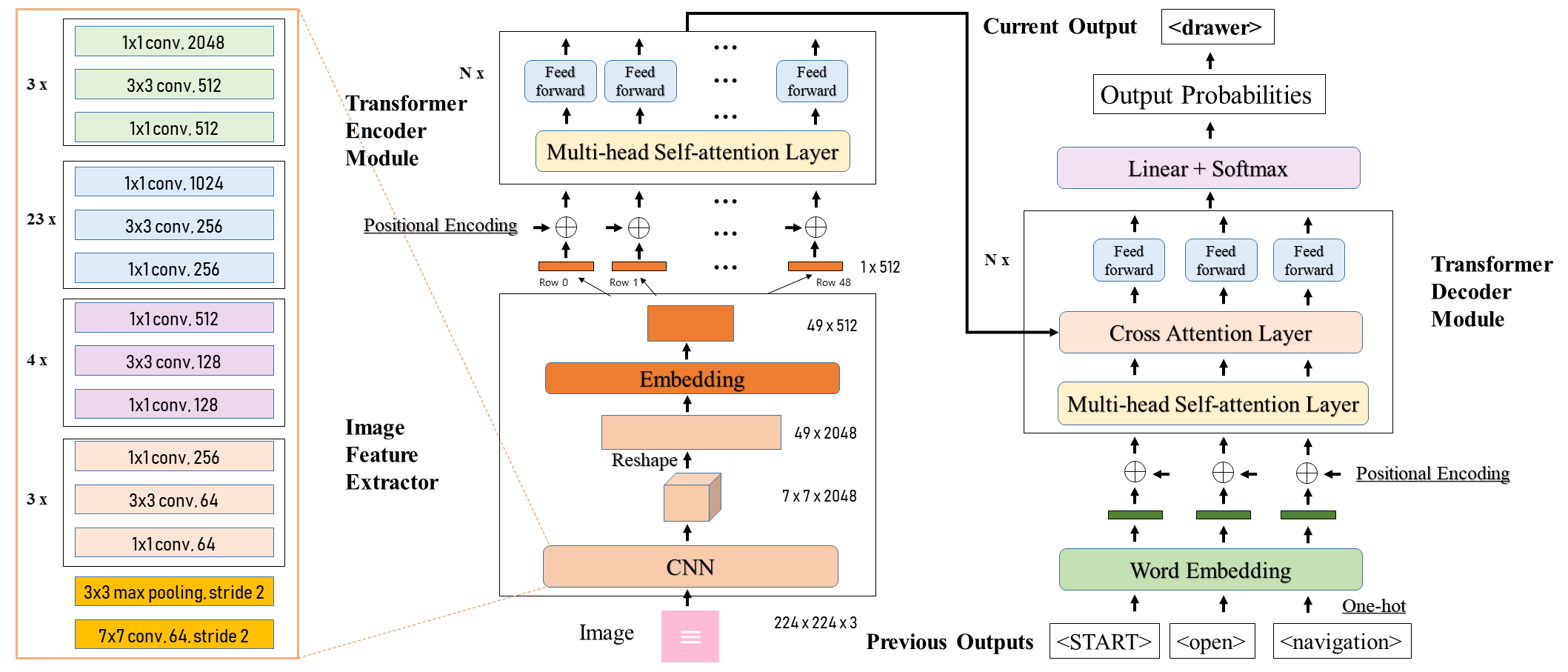}
	\caption{Overview of our approach }
	\label{fig:approach}
\end{figure*}

\section{Approach}
Rendering a UI widget into its corresponding content description is the typical task of image captioning.
The general process of our approach is to firstly extract image features using CNN~\cite{lecun1995convolutional}, and encode this extracted informative features into a tensor using an encoder module.
Based on the encoded information, the decoder module generates outputs (which is a sequence of words) conditioned on this tensor and previous outputs.
Different from the traditional image captioning methods based on CNN and RNN model or neural translation based on RNN encoder-decoder~\cite{chen2017community, gao2019neural, chen2019mining}, we adopt the Transformer model~\cite{vaswani2017attention} in this work.
The overview of our approach can be seen in Figure~\ref{fig:approach}.

\subsection{Visual Feature Extraction}
To extract the visual features from the input button image, we adopt the convolutional neural network~\cite{he2016deep} which is widely used in software engineering domain~\cite{chen2016learning, chen2018data, zhao2019actionnet}.
CNN-based model can automatically learn latent features from a large image database, which has outperformed hand-crafted features in many computer vision tasks~\cite{krizhevsky2012imagenet, ren2015faster}.
CNN mainly contains two kinds of layers, i.e., convolutional layers and pooling layers.

\textbf{Convolutional Layer.}
A Convolution layer performs a linear transformation of the input vector to extract and compress the salient information of input.
Normally, an image is comprised of a $H \times W \times C$ matrix where $H, W, C$ represent height, width, channel of this image respectively.
And each value in this matrix is in the range of 0 to 255.
A convolution operation uses several small trainable matrices (called kernels) to slide over the image along the width and height in a specified stride.
Each movement will compute a value of output at the corresponding position by calculating the sum of element-wise product of current kernel and current sub-matrix of the image.
One kernel performs one kind of feature extractor, and computes one layer of the final feature map.
For example, for an image $I\in R^{HWC}$, we feed it in a convolutional layer with stride one and $k$ kernels, the output will have the dimension of $H' \times W' \times k$, where $H'$ and $W'$ are the times of movement along height \& width.

\textbf{Pooling Layer.}
A pooling layer is used to down-sample the current input size to mitigate the computational complexity required to process the input by extracting dominant features, which are invariant to position and rotation, of input.
It uses a fixed length of window to slide the image with a fixed stride and summarises current scanned sub-region to one value.
Normally, the stride is same as the window's size so that the model could filter meaningless features while maintaining salient ones.
There are many kinds of strategy to summarise sub-region.
For example, for max-pooling, it takes the maximum value in the sub-region to summarise current patch.
For average pooling, it takes the average mean of all values in current patch as the output value.

\subsection{Visual Semantic Encoding}
\label{sec:encoder}
To further encode the visual features extracted from CNN, we first embed them using a fully-connected layer and then adopt the encoder based on the Transformer model which was first introduced for the machine translation task.
We select the Transformer model as our encoder and decoder due to two reasons.
First, it overcomes the challenge of long-term dependencies~\cite{hochreiter1997long}, since it concentrates on all relationships between any two input vectors.
Second, it supports parallel learning because it is not a sequential learning and all latent vectors could be computed at the same time, resulting in the shorter training time than RNN model.
Within the encoder, there are two kinds of sublayers: the multi-head self-attention layer and the position-wise feed forward layer.

\textbf{Multi-head Self-attention Layer}
Given a sequence of input vectors $X=[x_1, x_2, ..., x_n]^{T}(X \in R^{n \times d_{embed}})$, a self-attention layer first computes a set of query ($Q$), key ($K$), value ($V$) vectors ($Q/K\in R^{d_k \times n}, V \in R^{d_v \times n}$) and then calculates the scores for each position vector with vectors at all positions.
Note that image-based buttons are artificial images rendered by the compiler with the specified order i.e., from left to right, from top to bottom.
Therefore, we also consider the sequential spatial information to capture the dependency between the top-left and bottom-right features extracted by the CNN model.
For position $i$, the scores are computed by taking the dot product of query vector $q_i$ with all key vectors $k_j(j \in {1,2,...n})$.

In order to get a more stable gradient, the scores are then divided by $\sqrt{d_{k}}$.
After that, we apply a softmax operation to normalize all scores so that they are added up to 1.
The final output of the self-attention layer is to multiply each value vector to its corresponding softmax score and then sums it up.
The matrix formula of this procedure is:
\begin{equation}
Self\_Attention(Q, K, V) = softmax(\frac{QK^T}{\sqrt{d_k}})V
\end{equation}
where $Q = W_{q}^{e}X^T$, $K = W_{k}^{e}X^T$, $V= W_{v}^{e}X^T$ and $W_{q}^{e} \in R^{d_k \times d_{embed}}$, $W_{k}^{e} \in R^{d_k \times d_{embed}}$, $W_{v}^{e} \in R^{d_v \times d_{embed}}$ are the trainable weight metrics.
The $softmax(\frac{QK^T}{\sqrt{d_k}})$ can be regarded as how each feature ($Q$) of the feature sequence from CNN model is influenced by all other features in the sequence ($K$).
And the result is the weight to all features $V$.

The multi-head self-attention layer uses multiple sets of query, key, value vectors and computes multiple outputs.
It allows the model to learn different representation sub-spaces.
We then concatenate all the outputs and multiply it with matrix $W_o^{e} \in R^{d_{model} \times hd_v}$ (where h is the number of heads) to summarise the information of multi-head attention.

\textbf{Feed Forward Layer}
Then a position-wise feed forward layer is applied to each position.
The feed-forward layers have the same structure of two fully connected layers, but each layer is trained separately.
The feed forward layer is represented as:
\begin{equation}
Feed\_forward(Z) = W_2^{e} \times (W_1^{e} \times Z + b_1^{e}) + b_2^{e}
\end{equation}
where $W_1^{e} \in R^{d_{ff} \times d_{model}}$, $W_2^{e} \in R^{d_{model} \times d_{ff}}$, $b_1^{e} \in R^{d_{ff}}$, $b_2^{e} \in R^{ d_{model}}$ are the trainable parameters of two fully connected layers.

Besides, for each sub-layer, Transformer model applies residual connection~\cite{he2016deep} and layer normalization~\cite{ba2016layer}.
The equation is $LayerNorm(x + Sublayer(x))$, where x and Sublayer(x) are the input and output of current sub-layer.
For input embedding, Transformer model also applies position encoding to encode the relative position of the sequence.

\subsection{Content Description Generation}
As mentioned in Section~\ref{sec:encoder}, the encoder module is comprised of N stacks of layers and each layer consists of a multi-head self-attention sub-layer and a feed-forward layer.
Similar to the encoder, the decoder module is also of M-stack layers but with two main differences.
First, an additional cross-attention layer is inserted into the two sub-layers.
It takes the output of top-layer of the encoder module to compute query and key vectors so that it can help capture the relationship between inputs and targets.
Second, it is masked to the right in order to prevent attending future positions.

\textbf{Cross-attention Layer.}
Given current time $t$, max length of content description $L$, previous outputs
$ S^{d}(\in R^{d_{model} \times L}) $
from self-attention sub-layer and output $Z^{e}(\in R^{d_{model} \times n})$  from encoder module, the cross-attention sub-layer can be formulated as:
\begin{equation}
Cross\_Attention(Q^e, K^e, V^d) = softmax(\frac{Q^e(K^e)^T}{\sqrt{d_k}})V^d
\end{equation}
where
$Q^e = W^d_qZ$, $K^e = W^d_kZ$, $V^d = W^d_vS^{d}$,
$W^d_q \in R^{d_k \times d_{model}}$, $W^d_k \in R^{d_k \times d_{model}}$ and $ W^d_v \in R^{d_v \times d_{model}}$.
Note that we mask $S^{d}_k = 0$ (for $ k \geq t $) since we currently do not know future values.

\textbf{Final Projection.}
After that, we apply a linear softmax operation to the output of the top-layer of the decoder module to predict next word.
Given output $ D \in R^{d_{model} \times L}$ from decoder module, we have $Y' = Softmax(D*W_o^d + b_o^d)$ and take the $t_{th}$ output of Y' as the next predicted word.
During training, all words can be computed at the same time by masking $S^{d}_k = 0$ (for $ k \geq t $) with different $t$.
Note that while training, we compute the prediction based on the ground truth labels, i.e, for time t, the predicted word $y'_t$ is based on the ground truth sub-sequence $[y_0, y_1, ..., y_{t-1}]$.
In comparison, in the period of inference (validation/test), we compute words one by one, based on previous predicted words, i.e., for time t, the predicted word $y'_t$ is based on the ground truth sub-sequence $[y'_0, y'_1, ..., y'_{t-1}]$.

\textbf{Loss Function.}
We adopt Kullback-Leibler (KL) divergence loss~\cite{kullback1951information} (also called as relative entropy loss) to train our model.
It is a natural method to measure the difference between the generated probability distribution q and the reference probability distribution $p$.
Note that there is no difference between cross entropy loss and KL divergence since $D_{kl}(p|q) = H(p,q) - H(p)$, where H(p) is constant.

\begin{figure}
	\centering
	\fbox{\includegraphics[width=0.48\textwidth]{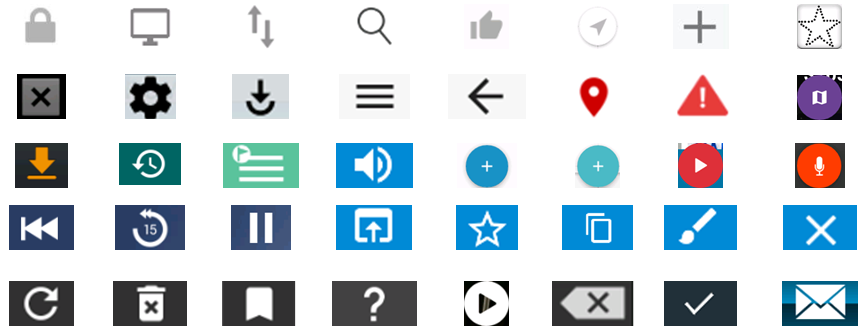}}
	\caption{Example of our dataset}
	\label{fig:dataset}
\end{figure}

\section{Implementation}
To implement our model, we further extract the data analysed in Section~\ref{sec:motivation} by filtering out the noisy data for constructing a large-scale pairs of image-based buttons and corresponding content descriptions for training and testing the model.
Then, we introduce the detailed parameters and training process of our model.

\subsection{Data Preprocessing}
\label{sec:data_preprocessing}
For each screenshot collected by our tool, we also dump the runtime XML which includes the position and type of all GUI components within the screenshot.
To obtain all image-based buttons, we crop the GUI screenshot by parsing the coordinates in the XML file.
However, the same GUI may be visited many times, and different GUIs within one app may share the same components.
For example, a menu icon may appear in the top of all GUI screenshots within the app.
To remove duplicates, we first remove all repeated GUI screenshots by checking if their corresponding XML files are the same.
After that, we further remove duplicate image-based buttons if they are exactly same by the pixel value.
But duplicate buttons may not be 100\% same in pixels, so we further remove duplicate buttons if their coordinate bounds and labels are the same because some buttons would appear in a fixed position but with a different background within the same app.
For example, for the ``back'' button at the top-left position in Figure~\ref{fig:illustration}, once user navigates to another news page, the background of this button will change while the functionality remains.

Apart from the duplicate image-based buttons, we also remove low-quality labels to ensure the quality of our training model.
We manually observe 500 randomly selected image-based buttons from Section~\ref{sec:motivation}, and summarise three types of meaningless labels.
First, the labels of some image-based buttons contain the class of elements such as ``image button'', ``button with image'', etc.
Second, the labels contain the app's name.
For example, the label of all buttons in the app \textsc{Ringtone Maker} is ``ringtone maker''.
Third, some labels may be some unfinished placeholders such as ``test'', ``content description'', ``untitled'', ``none''.
We write the rules to filter out all of them, and the full list of meaningless labels can be found in \oursite.

After removing the non-informative labels, we translate all non-English labels of image-based buttons to English by adopting the Google Translate API.
For each label, we add $<start>$, $<end>$ tokens to the start and the end of the sequence.
We also replace the low-frequency words with an $<unk>$ token.
To enable the mini-batch training, we need to add a $<pad>$ token to pad the word sequence of labels into a fixed length.
Note that the maximum number of words for one label is 15 in this work.

After the data cleaning, we finally collect totally 19,233 pairs of image-based buttons and content descriptions from 7,594 apps.
Note that the app number is smaller than that in Section~\ref{sec:motivation} as the apps with no or uninformative labels are removed.
We split cleaned dataset into training, validation\footnote{for tuning the hyperparameters and preventing the overfitting} and testing set.
For each app category, we randomly select 80\% apps for training, 10\% for validation and the rest 10\% for testing.
Table~\ref{tab:dataset_details} shows that, there are 15,595 image-based buttons from 6,175 apps as the training set, 1,759 buttons from 714 apps as validation set and 1,879 buttons from 705 apps as testing set.
The dataset can also be downloaded from our site.

\begin{table}
	\centering
	\caption{Details of our accessibility dataset.}
	\begin{tabular}{lrrr}
        \hline
            &\textbf{\#App} & \textbf{\#Screenshot}&  \textbf{\#Element}  \\
        \hline
        \textbf{Train}      & 6,175  & 10,566   & 15,595 \\
        \textbf{Validation} & 714    & 1,204    & 1,759 \\
        \textbf{Test}       & 705    & 1,375    & 1,879 \\
        \hline
        \textbf{Total}      & 7,594  & 13,145  & 19,233 \\
		\hline
	\end{tabular}
	\label{tab:dataset_details}
\end{table}

\subsection{Model Implementation}
We use ResNet-101 architecture~\cite{he2016deep} pretrained on ImageNet dataset~\cite{deng2009imagenet} as our CNN module.
As you can see in the leftmost of Figure~\ref{fig:approach}, it consists of a convolution layer, a max pooling layer, four types of blocks with different numbers of block (denoted in different colors).
Each type of block is comprised of three convolutional layers with different settings and implements an identity shortcut connection which is the core idea of ResNet.
Instead of approximating the target output of current block, it approximates the residual between current input and target output, and then the target output can be computed by adding the predicted residual and the original input vector.
This technique not only simplifies the training task, but also reduces the number of filters.
In our model, we remove the last global average pooling layer of ResNet-101 to compute a sequence of input for the consequent encoder-decoder model.

For transformer encoder-decoder, we take $N=3$, $d_{embed}=d_{k}=d_{v}=d_{model}=512$, $d_{ff}=2048$, $h=8$.
We train the CNN and the encoder-decoder model in an end-to-end manner using KL divergence loss~\cite{kullback1951information}.
We use Adam optimizer~\cite{kingma2014adam} with $\beta_1 = 0.9$, $\beta_2 = 0.98$ and $\epsilon = 10^{-9}$ and change the learning rate according to the formula $learning\_rate = d_{model}^{-0.5} \times min(step\_num^{-0.5}, step\_num \times warmup\_steps^{-1.5})$ to train the model,
where $step\_num$ is the current iteration number of training batch and the first $warm\_up$ training step is used to accelerate training process by increasing the learning rate at the early stage of training.
Our implementation uses PyTorch~\cite{web:pytorch} on a machine with Intel i7-7800X CPU, 64G RAM and NVIDIA GeForce GTX 1080 Ti GPU.

\begin{table*}
	\centering
	\caption{Results of accuracy evaluation}
	\begin{tabular}{|l|c|c|c|c|c|c|c|c|c|}
        \hline
        \textbf{Method}      &\textbf{Exact match} & \textbf{BLEU@1} & \textbf{BLEU@2} & \textbf{BLEU@3} & \textbf{BLEU@4} & \textbf{METEOR} & \textbf{ROUGE-L}   & \textbf{CIDEr}  \\
        \hline
        \textbf{CNN+LSTM}   & 58.4\% & 0.621 & 0.600 & 0.498 & 0.434 & 0.380 & 0.624 & 0.287 \\
        \textbf{CNN+CNN}    & 57.4\% & 0.618 & 0.596 & 0.506 & 0.473 & 0.374 & 0.617 & 0.284 \\
        \textbf{\tool } & \textbf{60.7\%} & \textbf{0.651} & \textbf{0.626} & \textbf{0.523} & \textbf{0.464} & \textbf{0.399} & \textbf{0.654} & \textbf{0.302} \\
		\hline
	\end{tabular}
	\label{tab:Accuracy Result}
\end{table*}

\section{Evaluation}

We evaluate our \tool in three aspects, i.e., accuracy with automated testing, generality and usefulness with user study.

\subsection{Evaluation Metric}
To evaluate the performance of our model, we adopt five widely-used evaluation metrics including exact match, BLEU~\cite{papineni2002bleu}, METEOR~\cite{banerjee2005meteor}, ROUGE~\cite{lin2003automatic}, CIDEr~\cite{vedantam2015cider} inspired by related works about image captioning.
The first metric we use is \textit{exact match rate}, i.e., the percentage of testing pairs whose predicted content description exactly matches the ground truth.
Exact match is a binary metric, i.e., 0 if any difference, otherwise 1.
It cannot tell the extent to which a generated content description differs from the ground-truth.
For example, the ground truth content description may contain 4 words, but no matter one or 4 differences between the prediction and ground truth, exact match will regard them as 0.
Therefore, we also adopt other metrics.
BLEU is an automatic evaluation metric widely used in machine translation studies.
It calculates the similarity of machine-generated translations and human-created reference translations (i.e., ground truth).
BLEU is defined as the product of $n$-gram precision and brevity penalty.
As most content descriptions for image-based buttons are short, we measure BLEU value by setting $n$ as 1, 2, 3, 4, represented as BLEU@1, BLEU@2, BLEU@3, BLEU@4.

METEOR~\cite{banerjee2005meteor} (Metric for Evaluation of Translation with Explicit ORdering) is another metric used for machine translation evaluation.
It is proposed to fix some disadvantages of BLEU which ignores the existence of synonyms and recall ratio.
ROUGE (Recall-Oriented Understudy for Gisting Evaluation) ~\cite{lin2003automatic} is a set of metric based on recall rate, and we use ROUGE-L, which calculates the similarity between predicted sentence and reference based on the longest common subsequence (short for LCS).
CIDEr (Consensus-Based Image Description Evaluation)~\cite{vedantam2015cider} uses term frequency inverse document frequency (tf-idf)~\cite{ramos2003using} to calculate the weights in reference sentence $s_{ij}$ for different n-gram $w_k$ because it is intuitive to believe that a rare n-grams would contain more information than a common one.
We use CIDEr-D, which additionally implements a length-based gaussian penalty and is more robust to gaming.
We then divide CIDEr-D by 10 to normalize the score into the range between 0 and 1.
We still refer CIDEr-D/10 to CIDEr for brevity.

All of these metrics give a real value with range [0,1] and are usually expressed as a percentage.
The higher the metric score, the more similar the machine-generated content description is to the ground truth.
If the predicted results exactly match the ground truth, the score of these metrics is 1 (100\%).
We compute these metrics using coco-caption code ~\cite{chen2015microsoft}.

\subsection{Baselines}
We set up two state-of-the-art methods which are widely used for image captioning as the baselines to compare with our content description generation method.
The first baseline is to adopt the CNN model to encode the visual features as the encoder and adopt a LSTM (long-short term memory unit) as the decoder for generating the content description~\cite{hochreiter1997long, vinyals2015show}.
The second baseline also adopt the encoder-decoder framework.
Although it adopts the CNN model as the encoder, but uses another CNN model for generating the output~\cite{aneja2018convolutional} as the decoder.
The output projection layer in the last CNN decoder performs a linear transformation and softmax, mapping the output vector to the dimension of vocabulary size and getting word probabilities.
Both methods take the same CNN encoder as ours, and also the same datasets for training, validation and testing.
We denote two baselines as \textit{CNN+LSTM}, \textit{CNN+CNN} for brevity.

\begin{table*}
	\centering
    \caption{Examples of wrong predictions in baselines}
	\scriptsize
	\setlength{\tabcolsep}{0.01em}
    \resizebox{\textwidth}{15mm}{
	\begin{tabular}{l Sc Sc Sc Sc Sc Sc Sc }
		\hline
		ID & E1 & E2 & E3 & E4  & E5  & E6 & E7  \\
		\hline
		\textbf{Button} & \fbox{\cincludegraphics[height=0.34cm]{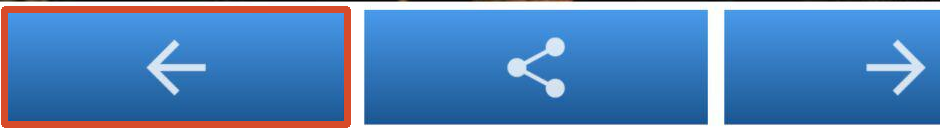}} & \fbox{\cincludegraphics[height=0.34cm]{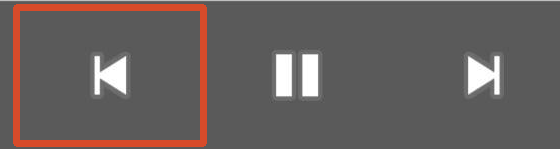}} &  \fbox{\cincludegraphics[height=0.345cm]{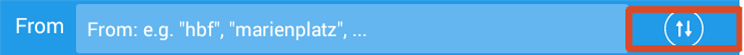}} & \fbox{\cincludegraphics[height=0.34cm]{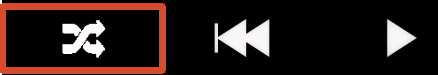}}& \fbox{\cincludegraphics[height=0.34cm]{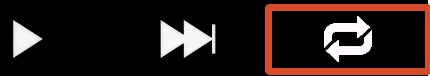}} & \fbox{\cincludegraphics[height=0.34cm]{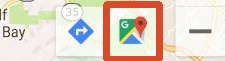}} & \fbox{\cincludegraphics[height=0.34cm]{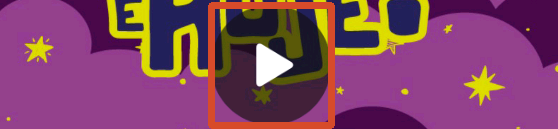}}  \\
		\hline
		\textbf{CNN+LSTM} & start  & play  & $<unk>$ & cycle shuffle mode & cycle repeat mode & color swatch & $<$unk$>$\\
		\textbf{CNN+CNN}  & next & previous track & call & cycle repeat mode   & next & open drawer  &  $<$unk$>$ trip check \\
		\textbf{\tool}  & back   & previous track  & \tabincell{c}{exchange origin and \\ destination points} & cycle shuffle mode  & cycle repeat mode& open in google maps  & watch\\
		\hline
	\end{tabular}
    }
	
	\label{tab:qualitative_example}
\end{table*}

\subsection{Accuracy Evaluation}
We use randomly selected 10\% apps including 1,879 image-based buttons as the test data for accuracy evaluation.
None of the test data appears in the model training.

\subsubsection{Overall Performance}
Table~\ref{tab:Accuracy Result} shows the overall performance of all methods.
The performance of two baselines are very similar and they achieve 58.4\% and 57.4\% exactly match rate respectively.
But CNN+LSTM model is slightly higher in other metrics.
In contrast with baselines, the generated labels from our \tool for 60.7\% image-based buttons exactly match the ground truth.
And the average BLEU@1, BLEU@2, BLEU@3, BLEU@4, METEOR, ROUGE-L and CIDEr of our method are 0.651, 0.626, 0.523, 0.464, 0.399, 0.654, 0.302.
Compared with the two state-of-the-art baselines, our \tool outperforms in all metrics and gains about 2\% to 11.3\% increase.
We conduct the Mann–Whitney U test~\cite{mann1947test} between these three models among all testing metrics.
Since we have three inferential statistical tests, we apply the Benjamini \& Hochberg (BH) method~\cite{benjamini1995controlling} to correct p-values.
Results show the improvement of our model is significant in all comparisons (p-value<0.01)\footnote{The detailed p-values are listed in \website}.

\begin{figure}
	\centering
    \includegraphics[width=0.5\textwidth]{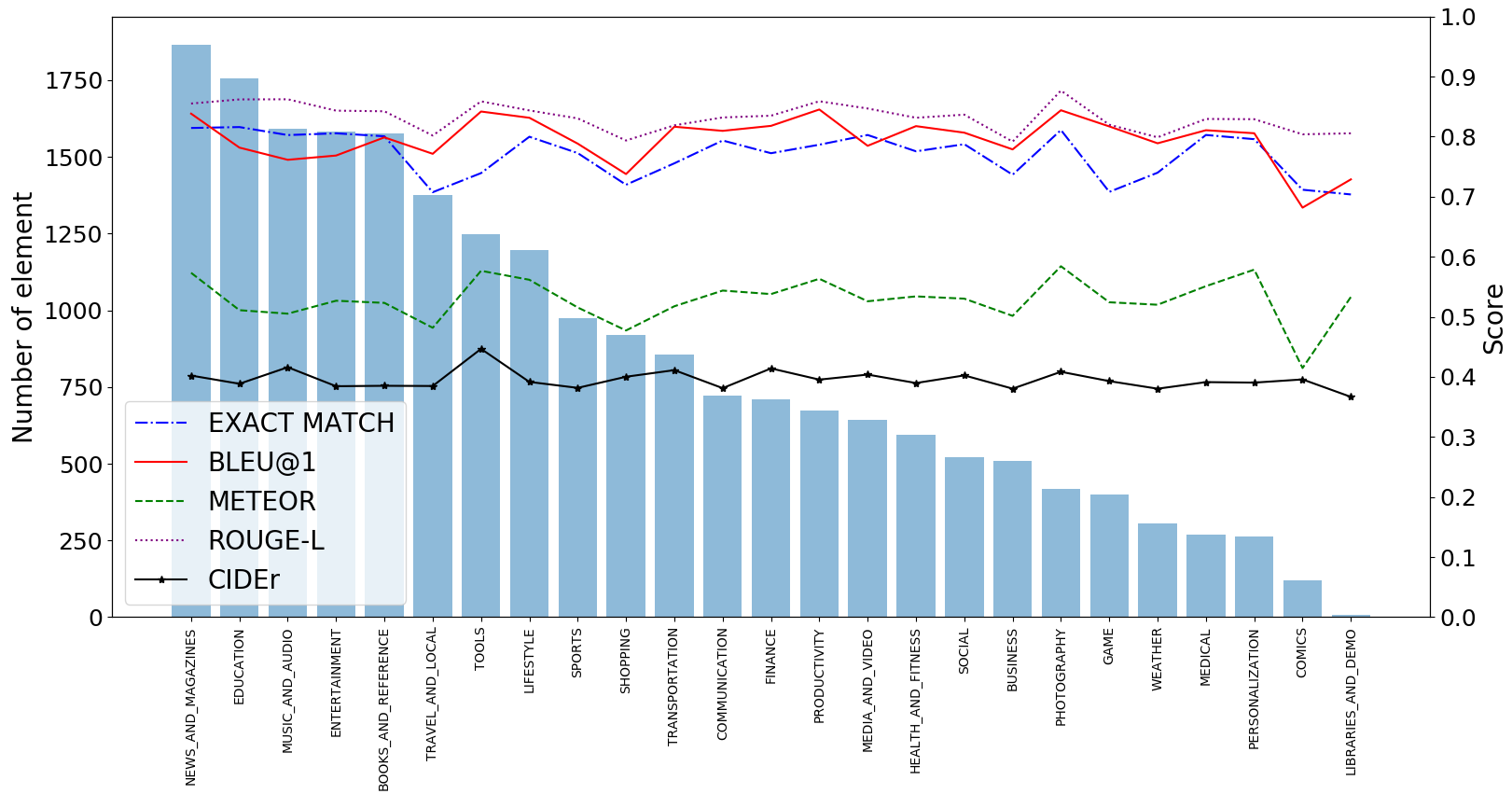}
	\caption{Performance distribution of different app category
    }
	\label{fig:train_cate_distribution}
\end{figure}

To show the generalization of our model, we also calculate the performance of our model in different app categories as seen in Figure~\ref{fig:train_cate_distribution}.
We find that our model is not sensitive to the app category i.e., steady performance across different app categories. 
In addition, Figure~\ref{fig:train_cate_distribution} also demonstrates the generalization of our model.
Even if there are very few apps in some categories (e.g., medical, personalization, libraries and demo) for training, the performance of our model is not significantly degraded in these categories.

\subsubsection{Qualitative Performance with Baselines}
To further explore why our model behaves better than other baselines, we analyze the image-buttons which are correctly predicted by our model, but wrongly predicted by baselines.
Some representative examples can be seen in Table~\ref{tab:qualitative_example} as the qualitative observation of all methods' performance.
In general, our method shows the capacity to generate different lengths of labels while CNN+LSTM prefers medium length labels and CNN+CNN tends to generate short labels.

Our model captures the fine-grained information inside the given image-based button.
For example, the CNN+CNN model wrongly predict ``next'' for the ``back'' button (E1), as the difference between ``back'' and ``next'' buttons is the arrow direction.
It also predicts ``next'' for the ``cycle repeat model'' button (E5), as there is one right-direction arrow.
Similar reasons also lead to mistakes in E2.

Our model is good at generating long-sequence labels due to the self-attention and cross-attention mechanisms.
Such mechanism can find the relationship between the patch of input image and the token of output label.
For example, our model can predict the correct labels such as ``exchange origin and destination points'' (E3), ``open in google maps'' (E6).
Although CNN+LSTM can generate correct labels for E4, E5, it does not work well for E3 and E6.

In addition, our model is more robust to the noise than the baselines.
For example, although there is ``noisy'' background in E7, our model can still successfully recognize the ``watch'' label for it.
In contrast, the CNN+LSTM and CNN+CNN are distracted by the colorful background information with ``$<$unk$>$'' as the output.

\subsubsection{Common Causes for Generation Errors}
We randomly sample 5\% of the wrongly generated labels for the image-based buttons.
We manually study the differences between these generated labels and their ground truth.
Our qualitative analysis identifies three common causes of the generation errors.

(1) Our model makes mistakes due to the characteristics of the input.
Some image-based buttons are very special and it is totally different from the training data.
For example, the E1 in Table~\ref{tab:common_causes} is rather different from the normal social-media sharing button.
It aggregates the icons of whatsapp, twitter, facebook and message with some rotation and overlap.
Some buttons are visually similar to others but with totally different labels.
The ``route'' button (E2) in Table~\ref{tab:common_causes} includes a right arrow which also frequently appears in ``next'' button.
(2) Our prediction is an alternative to the ground truth for certain image-based buttons.
For example, the label of E3
is the ``menu'', but our model predicts ``open navigation drawer''.
Although the prediction is totally different from the ground truth in term of the words, they convey the same meaning which can be understood by the blind users.
(3) A small amount of ground truth is not the right ground truth.
For example, some developers annotate the E4 as ``story content image'' although it is a ``download'' button.
Although our prediction is different from the ground truth, we believe that our generated label is more suitable for it.
This observation also indicates the potential of our model in identifying wrong/uninformative content description for image-based buttons.
We manually allocate the 5\% (98) failure cases of our model into these three reasons.
54 cases account for model errors especially for special cases, 41 cases are alternatives labels to ground truth, and the last three cases are right but with a wrong groundtruth.
It shows that the accuracy of our model is highly underestimated.

\begin{table}
	\centering
	\caption{Common causes for generation failure.}
    \tiny
    \setlength{\tabcolsep}{0.01em}
    \resizebox{0.5\textwidth}{15mm}{
	\begin{tabular}{l Sc Sc Sc Sc}
		\hline
		ID & E1 & E2 & E3 & E4  \\
		\hline
		\textbf{Cause} & Special case & Model error & Alternative & Wrong ground truth \\
		\hline
		\textbf{Button} & \fbox{\cincludegraphics[height=0.45cm]{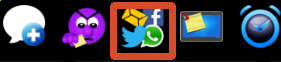}} & \fbox{\cincludegraphics[height=0.45cm]{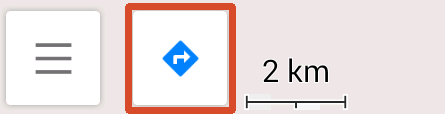}}   & \fbox{\cincludegraphics[height=0.45cm]{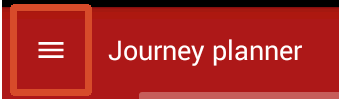}} & \fbox{\cincludegraphics[height=0.45cm]{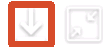}} \\
		\hline
		\textbf{\tool}  & $<unk>$ & next  & open navigation drawer & download \\
		\textbf{Ground truth}  & share note  & route & menu  & story content image  \\
		\hline
	\end{tabular}
    }
	\label{tab:common_causes}
\end{table}

\subsection{Generalization and Usefulness Evaluation}
To further confirm the generalization and usefulness of our model, we randomly select 12 apps in which there are missing labels of image-based buttons.
Therefore, all data of these apps do not appear in our training/testing data.
We would like to see the labeling quality from both our \tool and human developers.

\subsubsection{Procedures}
To ensure the representativeness of test data, 12 apps that we select have at least 1M installations (popular apps often influence more users), with at least 15 screenshots. 
These 12 apps belong to 10 categories.
We then crop elements from UI screenshots and filter out duplicates by comparing the raw pixels with all previous cropped elements.
Finally, we collect 156 missing-label image-based buttons, i.e., 13 buttons in average for each app.

All buttons are feed to our model for predicting their labels (denoted as M).
To compare the quality of labels from our \tool and human annotators, we recruit three PhD students and research staffs (denoted as A1, A2, A3) from our school to create the content descriptions for all image-based buttons.
All of them have at least one-year experience in Android app development, so they can be regarded as junior app developers.
Before the experiment, they are required to read the accessibility guidelines~\cite{web:androidAccessibityGuideline, web:talkbackGuideline} and we demo them the example labels for some randomly selected image-based buttons (not in our dataset).
During the experiment, they are shown the target image-based buttons highlighted in the whole UI (similar to Figure~\ref{fig:illustration}), and also the meta data about the app including the app name, app category, etc.
All participants carried out experiments independently without any discussions with each other.

As there is no ground truth for these buttons, we recruit one professional developer (evaluator) with prior experience in accessibility service during app development to manually check how good are the annotators' comments.
Instead of telling if the result is right or not, we specify a new evaluation metric for human evaluators called acceptability score according to the acceptability criterion~\cite{goto2013overview}.
Given one predicted content description for the button, the human evaluator will assign 5-point Likert scale~\cite{brooke1996sus, oda2015learning}) with 1 being least satisfied and 5 being most satisfied.
Each result from \tool and human annotators will be evaluated by the evaluator, and the final acceptability score for each app is the average score of all its image-based buttons.
Note that we do not tell the human evaluator which label is from developers or our model to avoid potential bias.
To guarantee if the human evaluators are capable and careful during the evaluation, we manually insert 4 cases which contain 2 intentional wrong labels and 2 suitable content description (not in 156 testing set) which are carefully created by all authors together.
After the experiment, we ask them to give some informal comments about the experiment, and we also briefly introduce \tool to developers and the evaluator and get some feedback from them.

\begin{figure}
	\centering
    \includegraphics[width=0.3\textwidth]{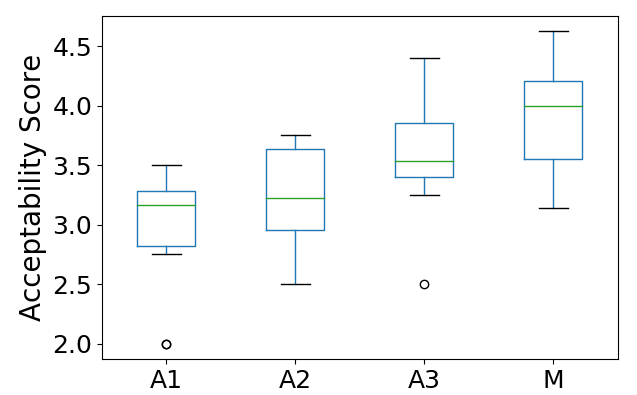}
	\caption{Distribution of app acceptability scores by human annotators (A1, A2, A3) and the model (M).
    }
	\label{fig:generalization_boxplot}
\end{figure}

\begin{table*}
    \centering
    \caption{The acceptability score (AS) and the standard deviation for 12 completely unseen apps. * denotes $p<0.05$.}
    \rowcolors{2}{gray!25}{white}
    \footnotesize
    \setlength{\tabcolsep}{0.8em}
    \resizebox{0.99\textwidth}{!}{
    \begin{tabular}{lllllllll}
    \hline
    ID & \textbf{Package name} & \textbf{Category} & \textbf{\#Installation} & \textbf{ \#Image-based button} & \textbf{AS-M} &  \textbf{AS-A1} & \textbf{AS-A2} & \textbf{AS-A3} \\
    \hline
    1 & com.handmark.sportcaster &sports & 5M - 10M &  8 & 4.63(0.48) & 3.13(0.78) & 3.75(1.20) & 4.38(0.99) \\

    2 & com.hola.launcher & personalization & 100M - 500M & 10 &4.40(0.92) & 3.20(1.08) & 3.50(1.75) & 3.40(1.56)  \\
    3 & com.realbyteapps.moneymanagerfree & finance & 1M - 5M  &  24 & 4.29(1.10) & 3.42(1.29)  & 3.75(1.45) & 3.83(1.55) \\
    4 & com.jiubang.browser & communication &  5M - 10M & 11 & 4.18(1.34) & 3.27(1.21) &  3.73(1.54) & 3.91(1.38) \\
    5 & audio.mp3.music.player & media\_and\_video & 5M - 10M &  26  & 4.08(1.24) & 2.85(1.06) & 2.81(1.62) & 3.50(1.62)\\
    6 & audio.mp3.mp3player    & music\_and\_audio &  1M - 5M &  16  & 4.00(1.27) & 2.75(1.15) & 3.31(1.53) & 3.25(1.39) \\
    7 & com.locon.housing      & lifestyle &  1M - 5M & 10 & 4.00(0.77) & 3.50(1.12) & 3.60(1.28) & 4.40(0.80) \\
    8 & com.gau.go.launcherex.gowidget.weatherwidget & weather & 50M - 100M & 12 & 3.42(1.66) & 2.92(1.38) & 3.00(1.78) & 3.42(1.80) \\
    9 & com.appxy.tinyscanner  & business & 1M - 5M &  13 & 3.85(1.23) & 3.31(1.20) & 3.08(1.59) & 3.38(1.44) \\
    10 & com.jobkorea.app      & business & 1M - 5M &  15 & 3.60(1.67) & 3.27(1.57) & 3.13(1.67) & 3.60(1.54) \\
    11 & browser4g.fast.internetwebexplorer & communication & 1M - 5M  &  4 & 3.25(1.79) & 2.00(0.71) & 2.50(1.12) & 2.50(1.66) \\

    12 & com.rcplus & social & 1M - 5M &  7 & 3.14(1.55) & 2.00(1.20) & 2.71(1.58) & 3.57(1.29) \\
    \hline
    \multicolumn{4}{c}{\textbf{AVERAGE}} & 13 & 3.97*(1.33) & 3.06(1.26) & 3.27(1.60) & 3.62(1.52) \\
    \hline
    \end{tabular}
    }

    \label{tab:generalization_results}
\end{table*}

\begin{table*}
	\centering
	\caption{Examples of generalization.}
	\resizebox{1.0\textwidth}{!}{
		\begin{tabular}{l Sc Sc Sc Sc Sc}
			\hline
			ID & E1 & E2 & E3 & E4 & E5  \\
			\hline
			\textbf{Button} & \fbox{\cincludegraphics[height=0.5cm]{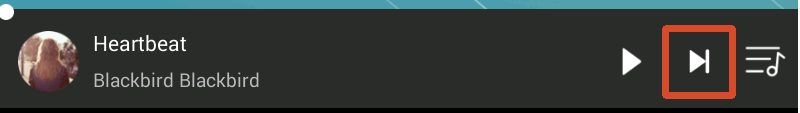}} & \fbox{\cincludegraphics[height=0.5cm]{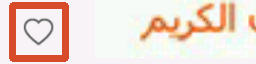}} & \fbox{\cincludegraphics[height=0.5cm]{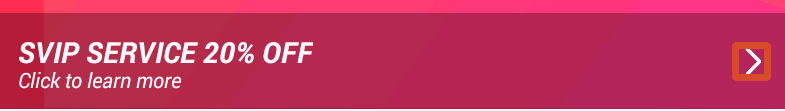}} & \fbox{\cincludegraphics[height=0.5cm]{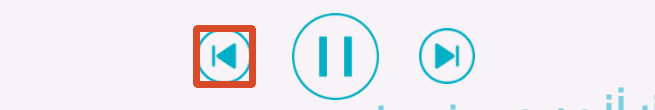}} & \fbox{\cincludegraphics[height=0.5cm]{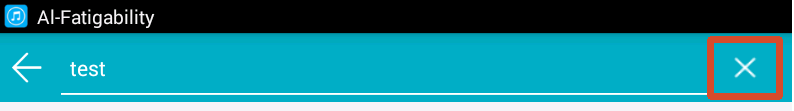}}  \\
			\hline
			\textbf{M}  & next song & add to favorites     &  open ad   & previous song & clear query  \\
			\textbf{A1} & change to the next song in playlist  & add the mp3 as favorite & show more details about SVIP& paly the former one & clean content \\
			\textbf{A2} & play the next song & like & check & play the last song& close  \\
			\textbf{A3} & next   &like & enter   & last & close \\
			\hline
		\end{tabular}
	}
	\label{tab:generalization_example}
\end{table*}

\subsubsection{Results}
Table~\ref{tab:generalization_results}\footnote{Detailed results are at \website} summarizes the information of the selected 12 apps and the acceptability scores of the generated labels.
The average acceptability scores for three developers vary much from 3.06 (A1) to 3.62 (A3).
But our model achieves 3.97 acceptability score which significantly outperforms three developers by 30.0\%, 21.6\%, 9.7\%.
The evaluator rates 51.3\% of labels generated from \tool as highly acceptable (5 point), as opposed to 18.59\%, 33.33\%, 44.23\% from three developers.
Figure~\ref{fig:generalization_boxplot} shows that our model behaves betters in most apps compared with three human annotators.
These results show that the quality of content description from our model is higher than that from junior Android app developers.
Note that the evaluator is reliable as both 2 intentional inserted wrong labels and 2 good labels get 1 and 5 acceptability score as expected.

To understand the significance of the differences between four kinds of content description, we carry out the Wilcoxon signed-rank test~\cite{wilcoxon1992individual} between the scores of our model and each annotator and between the scores of any two annotators.
It is the non-parametric version of the paired T-test and widely used to evaluate the difference between two related paired sample from the same probability distribution.
The test results suggest that the generated labels from our model are significantly better than that of developers ($p$-value $<$ 0.01 for A1, A2, and $<$ 0.05 for A3)\footnote{The p-values are adjusted by Benjamin \& Hochberg method~\cite{benjamini1995controlling}. All detailed p-values are listed in \website}.

For some buttons, the evaluator gives very low acceptability score to the labels from developers.
According to our observation, we summarise four reasons accounting for those bad cases and give some examples in Table~\ref{tab:generalization_example}.
(1) Some developers are prone to write long labels for image-based buttons like the developer A1.
Although the long label can fully describe the button (E1, E2), it is too verbose for blind users especially when there are many image-based buttons within one page.
(2) Some developers give too short labels which may not be informative enough for users.
For example, A2 and A3 annotate the ``add to favorite'' button as ``like'' (E2).
Since this button will trigger an additional action (add this song to favorite list), ``like'' could not express this meaning.
The same reason applies to  A2/A3's labels for E2 and such short labels do not contain enough information.
(3) Some manual labels may be ambiguous which may confuse users.
For example, A2 and A3 annotate ``play the last song'' or ``last'' to ``previous song'' button (E4) which may mislead users that clicking this button will come to the final song in the playlist.
(4) Developers may make mistakes especially when they are adding content descriptions to many buttons.
For example, A2/A3 use ``close'' to label a ``clear query'' buttons (E5).
We further manually check 135 low-quality (acceptability score = 1) labels from annotators into these four categories.
18 cases are verbose labels, 21 of them are uninformative, six cases are ambiguous which would confuse users, and the majority, 90 cases are wrong.

We also receive some informal feedback from the developers and the evaluator.
Some developers mention that one image-based button may have different labels in different context, but they are not very sure if the created labels from them are suitable or not.
Most of them never consider adding the labels to UI components during their app development and curious how the screen reader works for the app.
All of them are interested in our \tool and tell that the automatic generation of content descriptions for icons will definitely improve the user experience in using the screen reader.
All of these feedbacks indicate their unawareness of app accessibility and also confirm the value of our tool.

\section{Related Work}
Mobile devices are ubiquitous, and mobile apps are widely used for different tasks in people's daily life.
Consequently, there are many research works for ensuring the quality of mobile apps~\cite{butler2010android, fu2013people, hecht2015tracking}.
Most of them are investigating the apps' functional and non-functional properties like compatibility~\cite{wei2016taming},  performance~\cite{linares2015developers, zhao2016novel}, energy-efficiency~\cite{banerjee2016automated, banerjee2016debugging}, GUI design~\cite{chen2019storydroid, chen2019gallery}, GUI animation linting~\cite{zhao2020seenomaly}, localization~\cite{wang2019domain} and privacy and security~\cite{dehling2015exploring, yuan2016droiddetector, chen2019gui, chen2018automated, feng2019mobidroid}.
However, few of them are studying the accessibility issues, especially for users with vision impairment which is focused in our work.

\subsection{App Accessibility Guideline}
Google and Apple are the primary organizations that facilitate mobile technology and the app marketplace by Android and IOS platforms.
With the awareness of the need to create more accessible apps, both of them have released developer and designer guidelines for accessibility~\cite{web:androidAccessibityGuideline, web:iosGuidline} which include not only the accessibility design principles, but also the documents for using assistive technologies embedding in the operating system~\cite{web:iosGuidline}, and testing tools or suits for ensuring the app accessibility.
The World Wide Web Consortium (W3C) has released their web accessibility guideline long time ago~\cite{web:W3Access}
And now they are working towards adapting their web accessibility guideline~\cite{web:W3Access} by adding mobile characteristics into mobile platforms.
Although it is highly encouraged to follow these guidelines, they are often ignored by developers.
Different from these guidelines, our work is specific to users with vision impairment and predicts the label during the developing process without requiring developers to fully understand long guidelines.

\subsection{App Accessibility Studies for Blind Users}
Many works in Human-Computer Interaction area have explored the accessibility issues of small-scale mobile apps~\cite{yan2019current, park2014toward} in different categories such as in health~\cite{wang2012measurement}, smart cities~\cite{mora2017comprehensive} and government engagement~\cite{king2016government}.
Although they explore different accessibility issues, the lack of descriptions for image-based components has been commonly explicitly noted as a significant problem in these works.
Park et al~\cite{park2014toward} rated the severity of errors as well as frequency, and missing labels is rated as the highest severity of ten kinds of accessibility issues.
Kane et al~\cite{kane2009freedom} carry out a study of mobile device adoption and accessibility for people with visual and motor disabilities.
Ross et al~\cite{ross2018examining} examine the image-based button labeling in a relative large-scale android apps, and they specify some common labeling issues within the app.
Different from their works, our study includes not only the largest-scale analysis of image-based button labeling issues, but also a solution for solving those issues by a model to predict the label of the image.

There are also some works targeting at locating and solving the accessibility issues, especially for users with vision impairment.
Eler et al~\cite{eler2018automated} develop an automated test generation model to dynamically test the mobile apps.
Zhang et al~\cite{zhang2018robust} leverage the crowd source method to annotate the GUI element without the original content description.
For other accessibility issues, they further develop an approach to deploy the interaction proxies for runtime repair and enhancement of mobile application accessibility~\cite{zhang2017interaction} without referring to the source code.
Although these works can also help ensure the quality of mobile accessibility, they still need much effort from developers.
Instead, the model proposed in our work can automatically recommend the label for image-based components and developers can directly use it or modify it for their own apps.

\subsection{App Accessibility Testing Tools}
It is also worth mentioning some related non-academic projects.
There are mainly two strategies for testing app accessibility (for users with vision impairment) such as manual testing, and automated testing with analysis tools.
First, for manual testing, the developers can use the built-in screen readers (e.g., TalkBack~\cite{web:talkback} for Android, VoiceOver~\cite{web:voiceover} for IOS) to interact with their Android device without seeing the screen.
During that process, developers can find out if the spoken feedback for each element conveys its purpose.
Similarly, the Accessibility Scanner app~\cite{web:accessibilityScanner} scans the specified screen and provides suggestions to improve the accessibility of your app including content labels, clickable items, color contrast, etc.
The shortcoming of this tool is that the developers must run it in each screen of the app to get the results.
Such manual exploration of the application might not scale for larger apps or frequent testing, and developers may miss some functionalities or elements during the manual testing.

Second, developers can also automate accessibility tasks by resorting testing frameworks like Android Lint, Espresso and Robolectric, etc.
The Android Lint~\cite{web:androidLint} is a static tool for checking all files of an Android project, showing lint warnings for various accessibility issues including missing content descriptions and providing links to the places in the source code containing these issues.
Apart from the static-analysis tools, there are also testing frameworks such as Espresso~\cite{web:androidEspresso} and Robolectric~\cite{web:androidRobolectric} which can also check accessibility issues dynamically during the testing execution.
And there are counterparts for IOS apps like Earl-Grey~\cite{web:earlGrey} and KIF~\cite{web:kif}.
Note that all of these tools are based on official testing framework.
For example, Espresso, Robolectric and Accessibility Scanner are based on Android's Accessibility Testing Framework~\cite{web:accesstest}.

Although all of these tools are beneficial for the accessibility testing, there are still three problems
with them.
First, it requires developers' well awareness or knowledge of those tools, and understanding the necessity of accessibility testing.
Second, all these testing are reactive to existing accessibility issues which may have already harmed the users of the app before issues fixed.
In addition to these reactive testing, we also need a more proactive mechanism of accessibility assurance which could automatically predicts the content labeling and reminds the developers to fill them into the app.
The goal of our work is to develop a proactive content labeling model which can complement the reactive testing mechanism.

\section{Conclusion and Future Work}

More than 77\% apps have at least one image-based button without natural-language label which can be read for users with vision impairment.
Considering that most app designers and developers are of no vision issues, they may not understand how to write suitable labels.
To overcome this problem, we propose a deep learning model based on CNN and transformer encoder-decoder for learning to predict the label of given image-based buttons.

We hope that this work can invoking the community attention in app accessibility.
In the future, we will first improve our model for achieving better quality by taking the app metadata into the consideration.
Second, we will also try to test the quality of existing labels by checking if the description is concise and informative.

\bibliographystyle{ACM-Reference-Format}
\bibliography{reference}

\end{document}